# Strain-engineered high-temperature ferromagnetic Oxygen-substituted NaMnF$_3$ from first principles

Wen-ning Ren,[1,2] Kuijuan Jin,[1,2,3,*] Er-Jia Guo,[1,2] Chen Ge,[1,2] Can Wang,[1,2,3] Xiulai Xu,[1,2,3] Hongbao Yao,[1,2] Litong Jiang,[1,2] and Guozhen Yang[1,2]

[1]*Beijing National Laboratory for Condensed Matter Physics, Institute of Physics, Chinese Academy of Sciences, Beijing 100190, China*

[2]*School of Physical Sciences, University of Chinese Academy of Sciences, Beijing 100049, China*

[3]*Songshan Lake Materials Laboratory, Dongguan 523808, China*

[*]Author to whom correspondence should be addressed to Kuijuan Jin: kjjin@iphy.ac.cn

Using first-principles calculations, we investigated the magnetic, electronic, and structural properties of oxygen-substituted NaMnF$_3$ (NaMnF$_{1.5}$O$_{1.5}$) with in-plane biaxial strain. For simplicity, a structure containing an oxygen octahedron is used to explore the underlying physical mechanism. We found that the oxygen octahedron induces a transition from an insulating antiferromagnet to a high-temperature half-metallic ferromagnet. More importantly, the Curie temperature can be significantly enhanced and even might reach room temperature by applying tensile strain. The changing trends of exchange coupling constants with the increasing biaxial tensile strain can be attributed to the cooperative effects of Jahn-Teller distortion and rotation distortion. It is expected that these findings can enrich the versatility of NaMnF$_3$ and make it a promising candidate for spintronic applications.

**I INTRODUCTION**

Intriguing physical phenomena in perovskite materials have been of theoretical and experimental interest, e.g., magnetism [1], ferroelectricity [2,3], multiferroicity [4-6], piezoelectricity [7], and superconductivity [8]. Perovskite oxides are most extensively studied due to exotic physical properties harboring colossal magnetoresistance [9], unconventional superconductivity [10,11], which stems from a complex correlation among electronic, spin, structural, and orbital degrees of freedom [12,13]. In contrast, although some specific properties are demonstrated in halide perovskites [14] such as photocatalytic behavior in KMgF$_3$ [15] and the phase transition from the perovskite to the postperovskite in NaCoF$_3$ and NaNiF$_3$ [16,17], the researches on the magnetic properties for the fluoroperovskite are still scarce and desiderative [18,19].

NaMnF$_3$, an incipient multiferroic fluoroperovskite, having a coupling phenomenon of ferroelectricity and G-type antiferromagnetic (G-AFM) ordering with a transition temperature of 66 K from the antiferromagnetic to the paramagnetic state [6,20], has attracted significant interest theoretically [6,21] and experimentally [19,22,23]. Nevertheless, the antiferromagnetic ground state limits the applications of NaMnF$_3$ in information storage [24,25] and spintronics [26]. So far, the ferromagnetic property based on NaMnF$_3$ has never been investigated experimentally or theoretically, let alone the half-metallic ferromagnetism, which is one of the most demanded property for spintronics [27,28]. The previous studies have shown that, unlike oxides, in which the hybridization plays a dominant role, the effective ionic charges in fluoroperovskites are close to their nominal valence [6,21]. Furthermore, it is known that rotations and distortions of the oxygen octahedra affect the magnetic, electrical, and structural functionalities of perovskites [29-31]. Moreover, these properties can be conveniently tailored using external control parameters such as strain [32,33], electric field [34], and pressure [20]. For example, strain-induced magnetism in perovskite oxides was reported both theoretically [13,35] and experimentally [29,36]. Previous studies have also shown that the perovskite oxyfluorides, oxyhydrides, and oxynitrides have been synthesized by topochemical reactions, and novel effects of anion substitution on physical properties such as magnetic coupling [37,38] and dielectric properties [39] have been unveiled. For further exploring some novel property of the fluoroperovskite and establishing a link with perovskite oxides, here, by adopting the method of anion substitution, we theoretically predict a simple structure in which the oxygen octahedra are introduced into NaMnF$_3$ by replacing half fluorine atoms with oxygen atoms (NaMnF$_{1.5}$O$_{1.5}$), and its novel magnetic and transport properties, especially manipulated by the in-plane biaxial strain. We speculate that the structure is reasonable in accordance with the previous theoretical result [40], which means that NaMnF$_{1.5}$O$_{1.5}$ should exhibit well-ordered anion configuration due to the low-symmetry features inherited from precursor (NaMnF$_3$).

Except presenting the results of high-temperature ferromagnetism and electronic structures in NaMnF$_{1.5}$O$_{1.5}$ by employing first-principles calculations, we demonstrate that oxygen octahedron induces a transition from an insulating antiferromagnet to a late-model half-metallic ferromagnet and the Curie temperature ($T_C$) can be significantly increased by an in-plane biaxial tensile strain. Our results propose a new high-temperature half-metallic ferromagnet and illustrate the ability of strain engineering in tuning magnetism in NaMnF$_{1.5}$O$_{1.5}$.

## II COMPUTATIONAL METHOD

The first-principles calculations based on density functional theory (DFT) are performed by Vienna *ab initio* simulation package (VASP) [41,42]. The core and valence electrons are described by projected augmented wave (PAW) [43], with the electronic configuration as follows: 3s$^1$ (Na), 3$d^6$4$s^1$ (Mn), 2$s^2$2$p^5$ (F) and 2$s^2$2$p^6$ (O). We apply the generalized gradient approximation (GGA) exchange-correlation interaction within the revised Perdew-Burke-Ernzerhof method for solids (PBEsol) [44]. The GGA+U method [45] is adopted to improve the description of on-site Coulomb interactions to the Mn $d$ orbital with $U_{\text{eff}}$ = 4.0 eV ($U_{\text{eff}} = U-J$, where the Hund exchange parameter $J$ is set to 0 eV.) according to the previous studies [20,46]. Different $U_{\text{eff}}$ are also conducted for magnetic ground state (See Fig. S1 in Supplemental Material for details [47]). For structural optimization, the GGA+U is adopted within different magnetic states to perform a full relaxation, including FM, C-AFM, A-AFM, and G-AFM in both NaMnF$_3$ and NaMnF$_{1.5}$O$_{1.5}$. The Fig. S2 [47] shows the effect of $U_{\text{eff}}$ on the structural parameters. An 800 eV kinetic energy cutoff of the plane-wave basis set is used for all calculations. The Monkhorst-Pack K-mesh of 6×4×6 is used in the full Brillouin zone. The maximum convergence forces are optimized until less than 0.01 eV/Å, and the convergence criterion for the energy differences is set as 1×10$^{-6}$ eV. The crystal structures in the paper are drawn by VESTA Program [48].

## Ⅲ RESULTS AND DISCUSSION

The side and top views of the orthorhombic perovskite NaMnF$_3$ with the space group *Pnma* (No. 62) [49] are shown in Figs. 1(a) and 1(b), respectively, originating from a distortion of the cubic Pm$\bar{3}$m perovskite. The tilting of the purple MnF$_6$ octahedra of the $a^-b^+a^-$ type according to the Glazer's labeling [31], the deformation of octahedra, and the displacement of Na$^+$ cations result in the change of structural symmetry, indicating the coupling between tilting of MnF$_6$ and Na$^+$ displacement exists in the orthorhombic structure of NaMnF$_3$ [6]. We obtained the lattice parameters $a$ = 5.72 Å, $b$ = 7.96 Å, and $c$ = 5.49 Å from our DFT calculations, the corresponding accuracy is 99.46%, 99.41%, and 99%, respectively, compared with the experimental data [50]. Figures 1(c) and 1(d) show the side and top views of NaMnF$_{1.5}$O$_{1.5}$ with space group $P\bar{1}$ (No. 2), respectively, by substituting fluorine atoms with oxygen atoms and introducing the oxygen octahedra into NaMnF$_3$. The detailed results are listed in Table I, where the lattice constants of NaMnF$_{1.5}$O$_{1.5}$ are smaller in analogy to the intrinsic NaMnF$_3$ due to the role of hybridization in oxide perovskites [51]. Four possible structures based on the position of MnO$_6$ octahedron in NaMnF$_{1.5}$O$_{1.5}$ are shown in Fig. S3 [47], while no significant variations of the lattice parameters, energy difference for the different magnetic states, or total energy for the ground state are observed, some details are listed in Table. S1 [47]. The density of states (DOSs) of four structures are basically the same, as shown in Fig. S4 [47]. Then, the formation energy $E_{for}$ of NaMnF$_{1.5}$O$_{1.5}$ is calculated to study the thermodynamical stability [52,53], which can be evaluated by the following expression:

$$E_{for} = E(\text{NaMnF}_{1.5}\text{O}_{1.5}) - E(\text{NaMnF}_3) - \frac{3}{2}E(\text{O}) + \frac{3}{2}E(\text{F}), \qquad (1)$$

where $E(\text{NaMnF}_{1.5}\text{O}_{1.5})$ and $E(\text{NaMnF}_3)$ are the total energies of NaMnF$_{1.5}$O$_{1.5}$ and NaMnF$_3$, respectively. $E(\text{O}) = \frac{1}{2}E_{O_2}$ and $E(\text{F}) =$ $\frac{1}{2}E_{F_2}$, where the $E_{O_2}$ and $E_{F_2}$ denote the energies of oxygen and fluorine molecules, respectively. The negative formation energy -2.51 eV indicates the perovskite oxyfluoride is experimentally feasible [53-56].

Furthermore, the tolerance factor $t$ is usually used as a descriptor for the stability of perovskite structure [57,58]. For the perovskite oxyfluorides NaMnF$_{1.5}$O$_{1.5}$, it is defined as

$$t = \frac{r_A + r_C}{\sqrt{2}(r_B + r_C)}, \tag{2}$$

where $r_A$ (1.02 Å), $r_B$, and $r_C$ are the ionic radii of the Na$^+$, the arithmetic average radii of coexisting Mn$^{3+}$ (0.65 Å) and Mn$^{2+}$ (0.67 Å), and the weighted average of two anion radii, respectively. Here,

$$r_C = x \cdot r_{O^{2-}} + (1 - x) \cdot r_{F^-}, \tag{3}$$

with $x$ denoting the fraction of oxygen, $r_{O^{2-}}$=1.40 Å, and $r_{F^-}$=1.33 Å. We calculate the value for $t$ = 0.83, implying the perovskite structure can be maintained.

To determine the relative stability between different magnetic states, four possible spin orderings are considered: FM, C-AFM, A-AFM, and G-AFM (see Fig. 2). The energy difference between different magnetic configurations indicates a phase transition from G-AFM-state NaMnF$_3$ to FM-state NaMnF$_{1.5}$O$_{1.5}$, and the NaMnF$_{1.5}$O$_{1.5}$ possesses high magnetic moments, as shown in Table I.

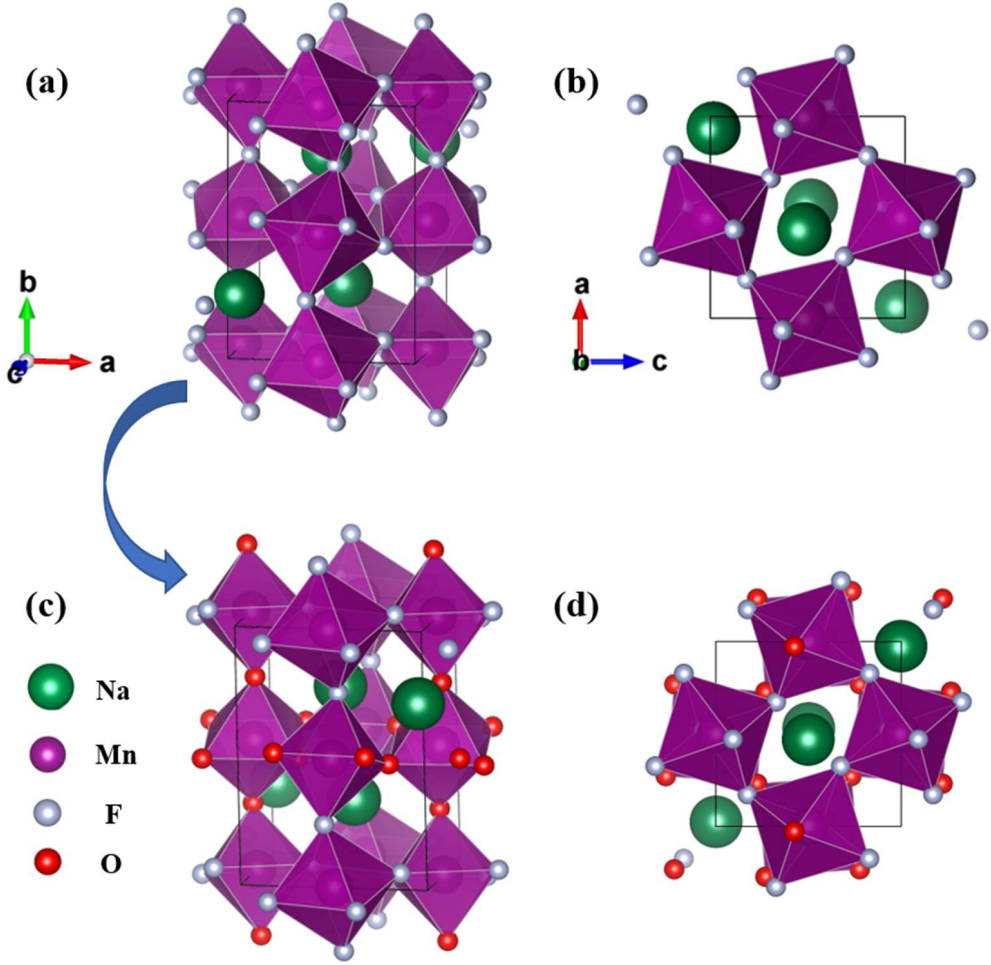

**FIG. 1.** Crystalline structures. (a) Side and (b) top views of orthorhombic perovskite NaMnF$_3$ (*Pnma*). (c) Side and (d) top views of NaMnF$_{1.5}$O$_{1.5}$. Green, purple, argenteous, and red balls represent Na, Mn, F, and O atoms, respectively. The corner-sharing octahedra are shown in purple.

TABLE I. The lattice constants, magnetic moments ($m^{Mn}$), energy difference ($\Delta E$), and magnetic exchange coupling parameters ($J_{ij}$) of NaMnF$_3$ and oxygen-substituted NaMnF$_3$ (NaMnF$_{1.5}$O$_{1.5}$). The Neel ($T_N$) and Curie temperatures ($T_C$) of NaMnF$_3$ and NaMnF$_{1.5}$O$_{1.5}$ are listed, respectively. The Mn1, Mn2, Mn3, and Mn4 denote four locations of Mn atoms in NaMnF$_3$, as well as in NaMnF$_{1.5}$O$_{1.5}$ crystal cell. The A, C, G, and F denote the total energy of A-type, C-type, G-type antiferromagnetic (AFM), and ferromagnetic (FM), respectively.

| Composition | Lattice constants (Å) | | $m^{Mn}$ ($\mu_B$/at) | | $\Delta E$ (meV) | | $J_{ij}$ (meV) | | $T$/(K) |
|---|---|---|---|---|---|---|---|---|---|
| NaMnF$_3$ | a | 5.72 | Mn1 | -4.675 | A-G | 54.631 | $J_1$ | -6.88 | |
| | b | 7.96 | Mn2 | 4.675 | C-G | 23.097 | $J_2$ | −5.88 | $T_N$ (151) |
| | c | 5.49 | Mn3 | 4.675 | F-G | 78.611 | $J_3$ | -0.03 | |
| | | | Mn4 | -4.675 | | | | | |
| NaMnF$_{1.5}$O$_{1.5}$ | a | 5.37 | Mn1 | 3.298 | A-F | 169.977 | $J_1'$ | 18.46 | |
| | b | 7.54 | Mn2 | 3.338 | C-F | 159.366 | $J_1''$ | 39.58 | $T_C$ (613) |
| | c | 5.35 | Mn3 | 4.476 | G-F | 306.014 | $J_1'''$ | 0.73 | |
| | | | Mn4 | 3.910 | | - | | | |

As discussed above from Table I, the introduction of oxygen octahedra can regulate the magnetic ground state of NaMnF$_3$ effectively. Therefore, we first estimate magnetic exchange coupling parameters by utilizing an effective Heisenberg Hamiltonian [59,60]:

$$H = E_0 - \sum_{i \neq j} J_{ij} e_i \cdot e_j, \tag{4}$$

where $E_0$ denotes the nonmagnetic ground-state energy, $e_i$ ($e_j$) denotes the unit vector along the direction of the magnetization at site $i$ ($j$), and $J_{ij}$ are exchange coupling parameters. The considered spin orderings for NaMnF$_3$ and NaMnF$_{1.5}$O$_{1.5}$ are depicted in Fig. 2. Here, the nearest-neighbor (NN) interaction parameter $J_1$, the next-nearest-neighbor (NNN) interaction parameter $J_2$, and the third-nearest-neighbor (3-NN) interaction parameter $J_3$ are taken into account for NaMnF$_3$, as shown in Fig. 2(a). Based on the Eq. (4), the linear equations can be expressed as follows:

$$E_{FM} = E_0 - 4J_1 - 2J_2 - 8J_3$$

$$E_{C-AFM} = E_0 + 4J_1 - 2J_2 + 8J_3$$

$$E_{A-A} = E_0 - 4J_1 + 2J_2 + 8J_3$$

$$E_{G-A} = E_0 + 4J_1 + 2J_2 - 8J_3. \tag{5}$$

The antiferromagnetic transition temperature $T_N$ can be estimated by mean-field approximation (MFA) [59,61]:

$$T_N = \frac{2}{3k_B}(-2J_1 - J_2 + 4J_3), \tag{6}$$

where $k_B$ is the Boltzmann constant. From Eq. (5) and Eq. (6), the calculated results of $J_{ij}$ are also listed in Table I, and $T_N = 151\ K$. The negative $J_1$, $J_2$, and $J_3$ indicate AFM interaction. Moreover, $J_1$ and $J_2$ are about 30 times larger than $J_3$, demonstrating that $J_1$ and $J_2$ play a major role and thus results in G-AFM ground state. Given that the MFA overestimates the transition temperature due to the neglected effect of spin fluctuation [62], the 151 K of $T_N$ is judged to be in quite good agreement with the experimental result [50,63].

We then estimate the Curie temperature $T_C$ of the FM-state NaMnF$_{1.5}$O$_{1.5}$. As the environments around Mn1, Mn2, Mn3, and Mn4 are different due to the anion substitution, corresponding to the MnF$_2$O$_4$, MnO$_6$, MnF$_6$, and MnF$_4$O$_2$ octahedra, respectively, in NaMnF$_{1.5}$O$_{1.5}$, for simplicity, three NN interactions between Mn atoms of different octahedra are considered as $J_1'$ (Mn1 and Mn2), $J_1''$ (Mn2 and Mn4), and $J_1'''$ (Mn2 and Mn3), respectively, as shown in Fig. 2(e). According to the Eq. (4), the linear equations are given as follows:

$$E_{FM} = E_0 - 4J_1' - 2J_1'' - 8J_1'''$$

$$E_{\text{C-AFM}} = E_0 + 4J_1' - 2J_1'' + 8J_1'''$$

$$E_{\text{A-A}} = E_0 - 4J_1' + 2J_1'' + 8J_1'''$$

$$E_{\text{G-AFM}} = E_0 + 4J_1' + 2J_1'' - 8J_1''' , \qquad (7)$$

and similarly, the $T_C$ can be estimated by MFA [59,61]:

$$T_C = \frac{2}{3k_B}(2J_1' + J_1'' - 4J_1''') . \qquad (8)$$

The calculation results indicate that $J_1''$ plays a crucial role in the high-temperature FM NaMnF$_{1.5}$O$_{1.5}$ (see Table I).

To reveal the mechanism of the magnetic transition from AFM NaMnF$_3$ to FM NaMnF$_{1.5}$O$_{1.5}$, the projected density of states (PDOS), indicating stronger hybridization between O and Mn atoms in NaMnF$_{1.5}$O$_{1.5}$, are plotted in Fig. S5(a) and S5(b), and the spin configurations of Mn $3d$ orbitals in NaMnF$_{1.5}$O$_{1.5}$ are also shown in Fig. S5(c). Compared with the pure Mn$^{2+}$ in NaMnF$_3$, some of the Mn$^{2+}$ are converted into Mn$^{3+}$ in NaMnF$_{1.5}$O$_{1.5}$, resulting in the hybrid $e_g$ of Mn$^{2+}$ interacts with the empty orbital $e_g$ of NN Mn$^{3+}$ through O$_{2p}$ orbital with a positive exchange coupling parameter. This double-exchange interaction [64] should be the origin for the FM feature of NaMnF$_{1.5}$O$_{1.5}$.

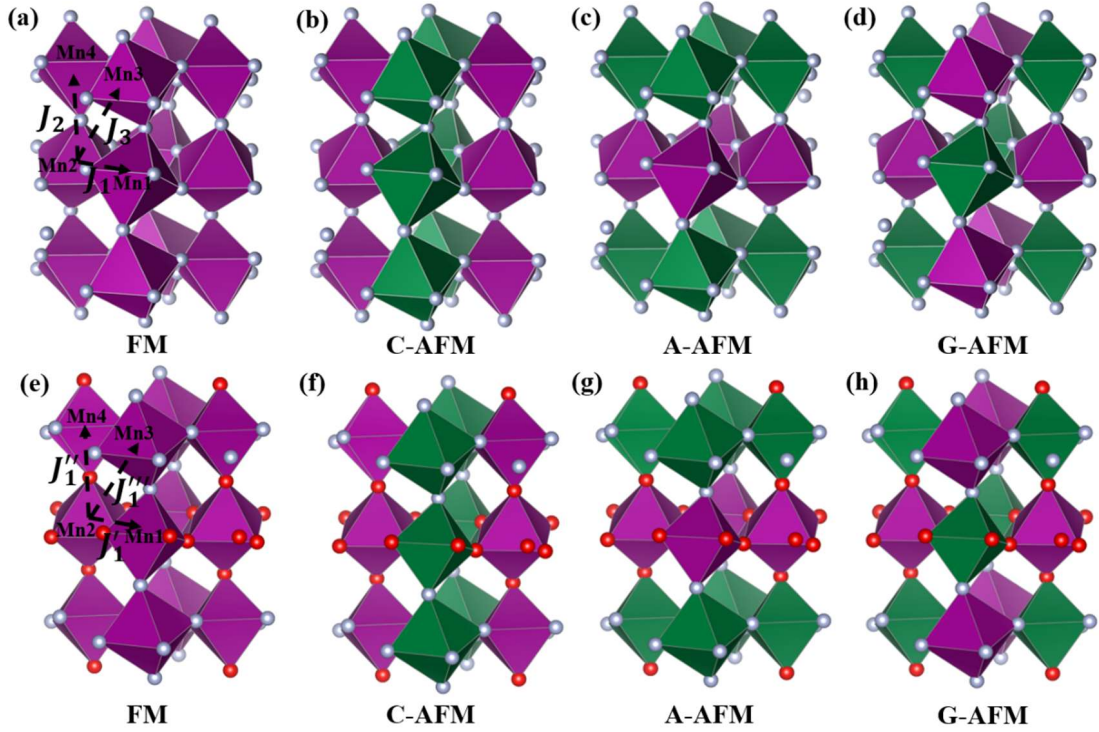

**FIG. 2.** Four considered magnetic configurations: (a) FM, (b) C-AFM, (c) A-AFM, (d) G-AFM of NaMnF$_3$. Four considered magnetic configurations: (e) FM, (f) C-AFM, (g) A-AFM, (h) G-AFM of NaMnF$_{1.5}$O$_{1.5}$. The purple and green octahedra denote up and down spins of magnetic Mn atoms, respectively. The Mn1, Mn2, Mn3, and Mn4 denote four locations of Mn atoms in NaMnF$_3$, as well as in NaMnF$_{1.5}$O$_{1.5}$ crystal cell. The exchange paths are indicated by black dashed arrows.

Strain engineering, as a promising way in manipulating the magnetism of transition metal compounds [46,65], providing an effective avenue to discover the novel magnetism of the ferromagnetic NaMnF$_{1.5}$O$_{1.5}$. Here, we define the strain as $\eta = \frac{d-d_0}{d_0}$, where $d_0$ and $d$ correspond to the pristine and strained in-plane lattice constants, respectively. Thus, the positive value represents tensile strain and negative value represents compressive strain. Herein, large strains are taken into account based on previous experimental method [66]. Due to the lowering effect of compressive strain on $T_C$ (see Fig. S6 [47]), we only discuss the effect of tensile strain on the NaMnF$_{1.5}$O$_{1.5}$, within which structural stability can be kept (see Fig. S7 [47]). The calculations of noncollinear magnetism are performed to test whether the magnetic ordering orientations change under the tensile strain (see Fig. S8 [47]). By applying different in-plane biaxial tensile strains to the NaMnF$_{1.5}$O$_{1.5}$, we predict the changing trend of exchange coupling constants $J_{ij}$, as shown in Fig. 3(a). The result clearly shows that the exchange coupling constants $J_1'$, $J_1''$, and $J_1'''$ increase with the increase in biaxial tensile strain from 0 to 6%. However, as the larger tensile strain than 6% is applied, $J_1'$ and $J_1'''$ show an obvious decay, and even the sign of $J_1'''$ changes from positive to negative, which means the occurrence of the transition from FM interaction to AFM interaction. As can be seen from Fig. 3(a), $J_1''$ plays a major role in the magnetic interactions. Fig. 3(b) shows that the $T_C$ can be strongly enhanced by the in-plane biaxial tensile strain from 613 K to 906 K in the strain range of 0 to 8%. As mentioned above, the MFA usually overestimates the transition temperature due to the neglected effect of spin fluctuation [62], and our calculated result $T_N$ (151 K) presented above is 2.3 times of data from the experimental measurement [50,63]. This kind of correlation coefficient between the theoretical and experimental values was also used by the previous researches [61,67,68]. Thus, it can be expected that a room temperature $T_C$ might be achieved with tensile strain less than 8%. With larger tensile strain, the $T_C$ shows a slow uptrend, which is the same trend as the major $J_1''$.

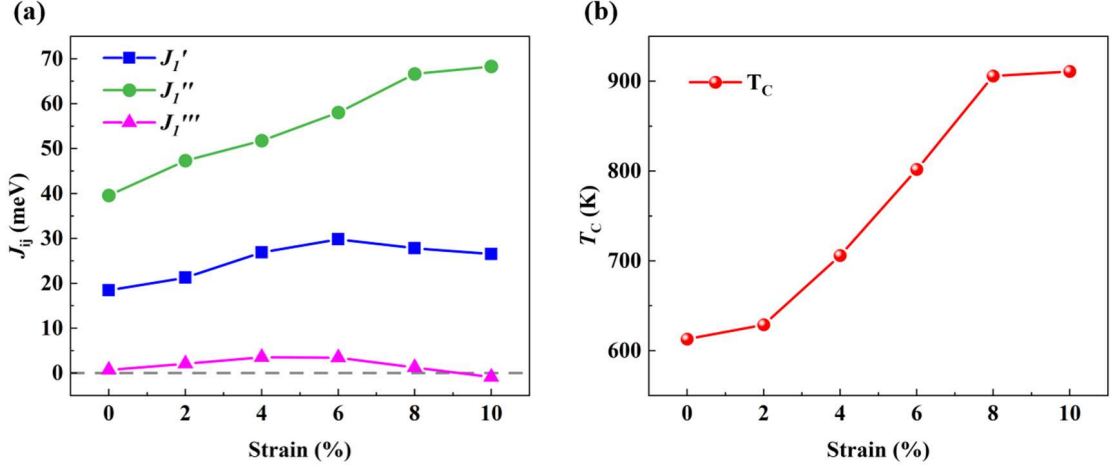

**FIG. 3.** (a) In-plane biaxial tensile strain dependencies of the magnetic-exchange interactions ($J_{ij}$). The gray horizontal dashed line denotes the boundary of magnetic interaction. (b) Evolution of Curie temperature ($T_C$) as a function of in-plane biaxial tensile strain.

To study the effect of in-plane biaxial tensile strain on structural distortions, we mainly focus on the oxygen-dominated octahedra, as shown in Fig. 4(a). It is generally believed that structural distortion of orthorhombic LaMnO$_3$-like perovskite consists of octahedral rotation distortion and Jahn-Teller (JT) distortion, which can be evaluated from the deviation of magnetic-ion-ligand-magnetic-ion bond angles from 180° and the relative size of Mn-O bonds in oxygen octahedra, respectively. [69]. Figure 4(a) shows the in-plane Mn-O-Mn bond angles θ$_1$ and θ$_2$, the out-of-plane Mn-O-Mn bond angle θ$_3$, and the oxygen-dominated octahedra. As shown in Fig. 4(b), the θ$_1$ increases as the tensile strain increases, while $\theta_2$ reduces when the applied tensile strain exceeds 6%, thus on the whole, in-plane Mn-O-Mn bond angles increase with the increase of tensile strain, which means in-plane rotation distortion is suppressed. The $\theta_3$ decreases as the tensile strain increases, indicating the out-of-plane rotation distortion is strengthened. Obvious changes in Mn-O-Mn bond angles indicate that rotation distortion of the oxygen-dominated octahedra contribute to the change in exchange interactions. The JT distortion, as an important role related to the structural distortions, is decomposed into the two normal modes: $Q_2$ and $Q_3$. In order to evaluate the strength of each distortion mode, we define the $Q_2 = \frac{l_x - l_y}{\sqrt{2}}$ and $Q_3 = \frac{2l_z - l_x - l_y}{\sqrt{6}}$, where the $l_x$, $l_y$, and $l_z$ denote the length of different Mn-O (Mn-F) bonds in oxygen-dominated octahedra, as shown in Fig. 4(a). Figures 4(c) and 4(d) show that the variation of $Q_2$ and $Q_3$ with tensile strain in fluorine-oxygen octahedron and oxygen octahedron, respectively. For the $Q_2$-type JT distortion, it remains almost unchanged with the increase of tensile strain in both fluorine-oxygen octahedron and oxygen octahedron. While the $Q_3$-type JT distortion is gradually enhanced with the increasing tensile strain in both octahedra. It should be noted that the value of $Q_3$ mode in an oxygen octahedron is about three times larger than that in the fluorine-oxygen octahedron, which suggests that the $Q_3$-type JT distortion in the oxygen octahedron plays a major role in the change of exchange interactions. We attribute the change of exchange interactions to the cooperative effect of rotation distortion and JT distortion.

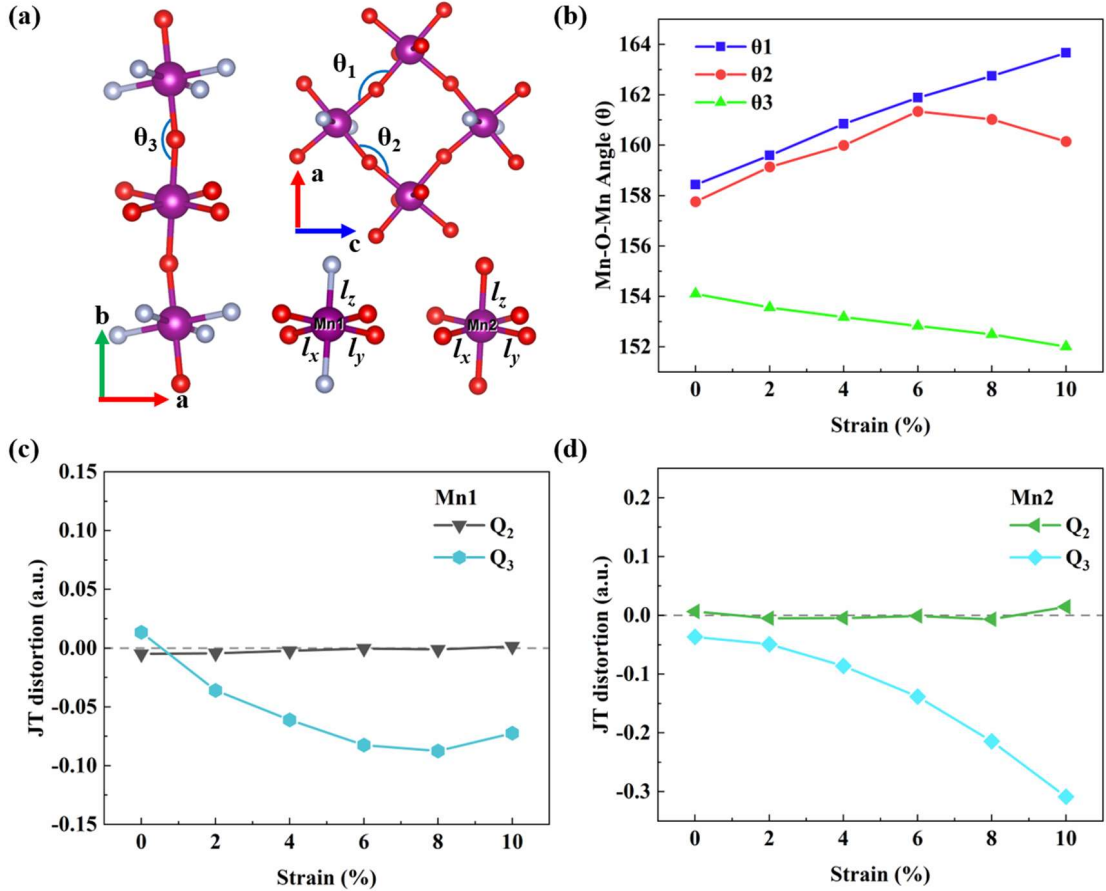

**FIG. 4.** (a) The Mn-O-Mn bond angles $\theta_1$ and $\theta_2$ from top view and $\theta_3$ along *b*-axis. For the oxygen-dominated octahedra, the in-plane Mn-O bonds are denoted by $l_x$ and $l_y$, and out-of-plane Mn-O bond is denoted by $l_z$. (b) Mn-O-Mn bond angles as a function of biaxial tensile strain. Evolution of the normal Jahn-Teller (JT) distortion mode (c) $Q_2$ and (d) $Q_3$.

To understand the effect of biaxial tensile strain on the changing trends of exchange interactions, we calculated the total density of states (TDOS) of NaMnF$_{1.5}$O$_{1.5}$ at zero strain and biaxial tensile strain of 10%, as shown in Fig. 5(a) and Fig. 5(b), respectively. One notable feature is that the Fermi level only across the spin-up states without any strain, which means that the spin-up channel possesses metallic conductivity while the spin-down channel is semiconducting, indicating a half-metallic character, as shown in Fig. 5(a). Moreover, Fig. 5(a) shows that the spin-flip excitation energy ($\Delta_{sf}$) is equal to the spin-down gap ($\Delta_\downarrow$) at zero strain. When a large tensile strain ($\eta =10\%$) is applied, the $\Delta_{sf}$ and $\Delta_\downarrow$ increase from 1.91 eV to 2.72 eV and 2.84 eV, respectively, as shown in Fig. 5(b). The enhanced spin polarization increases the spin transport distance, which indicates that the strain-tuned NaMnF$_{1.5}$O$_{1.5}$ is a potential candidate for spintronic materials. As shown in Fig. 5(c), the schematic electronic structures restructuring illustrates the regulation mechanism more visually.

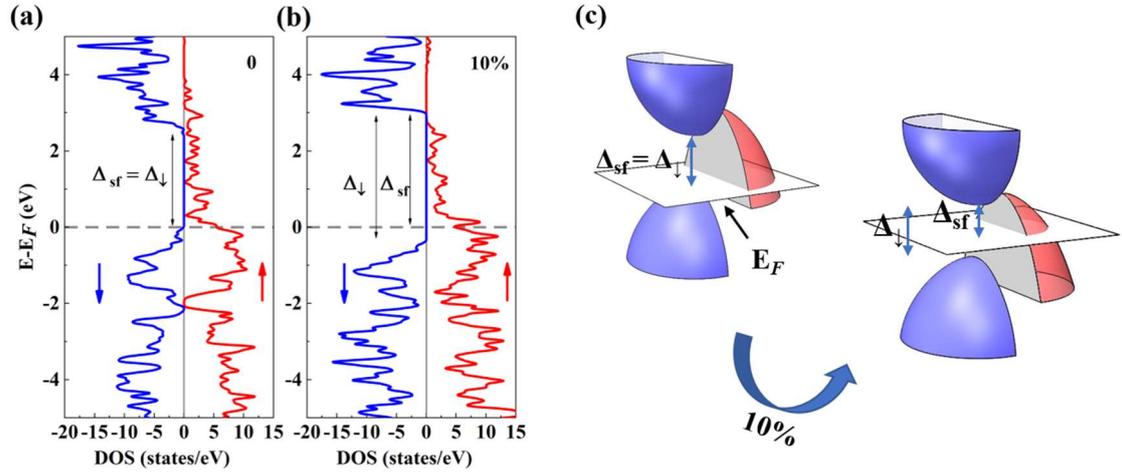

**FIG. 5.** The total density of states (TDOS) of NaMnF$_{1.5}$O$_{1.5}$ at (a) 0 strain and (b) biaxial tensile strain of 10%. The symbols $\Delta_{sf}$ and $\Delta_\downarrow$ represent the spin-flip excitation energy and spin-down gap, respectively. The Fermi level is set at zero. (c) The schematic of the electronic structures restructuring models at zero strain and biaxial tensile strain of 10%. The white plane denotes the Fermi surface. Spin-up and spin-down states are marked by red and blue, respectively.

We then investigated the PDOS on the Mn-3$d$ orbitals in MnO$_6$ octahedra (Mn2) at various biaxial tensile strain, as shown in Figs. 6(a)-6(f). For comparison, Fig. S9 [47] shows that the $t_{2g}$ orbitals change slightly with the increased biaxial tensile strain. According to the previous study [65,70], the $e_g$ ($d_{z^2}/d_{x^2-y^2}$) exchange interactions are FM and mainly determine the interplanar exchange. As mentioned above, the $Q_3$-type JT distortion of the oxygen octahedra gradually promotes as the applied biaxial tensile strain increases (see Fig. 4(d)), which enhanced the crystal field splitting. Further results can be seen from Figs. 6(a)-6(f), the tensile strain increases the orbital overlap between $d_{z^2}$ and $d_{x^2-y^2}$, implying the FM $e_g$-$e_g$ exchange interaction is enhanced, which explains the increase of the interplanar exchange coupling constant $J_1''$ with increasing biaxial tensile strain (see Fig. 3(a)). Furthermore, the bond-angle-related rotation distortion is considered. According to the Goodenough-Kanamori-Anderson rules, the super-exchange interaction favors AFM with 180° Mn-O-Mn bond angle [71-73]. The out-of-plane Mn-O-Mn bond angle θ$_3$ decreases monotonically with the increase of the biaxial tensile strains, as shown in Fig. 4(b), indicating the AFM interaction lessens, which also links the rotation distortion to the changing trend of the major exchange coupling constant $J_1''$ (see Fig. 3(a)). However, previous study confirmed that the $d_{z^2}$-$d_{z^2}$ FM exchange determine the changing trend of the intraplane exchange and is strongly affected by the in-plane Mn-O-Mn bond angle [65]. As shown in Figs. 6(a)-6(c), the intensity of the projected orbital $d_{z^2}$ changes slightly at the smaller biaxial tensile strain (0-4%). As the tensile strain increases, in-plane Mn-O-Mn bond angle $\theta_2$ gradually decreases and the intensity of the projected orbital $d_{z^2}$ increases obviously, indicating that the bandwidth of the $d_{z^2}$ decreases and the in-plane orbital overlaps between the $d_{z^2}$ states reduce. These results further lead to the reduction of the in-plane FM exchange coupling constant $J_1'$ (see Fig. 3(a)).

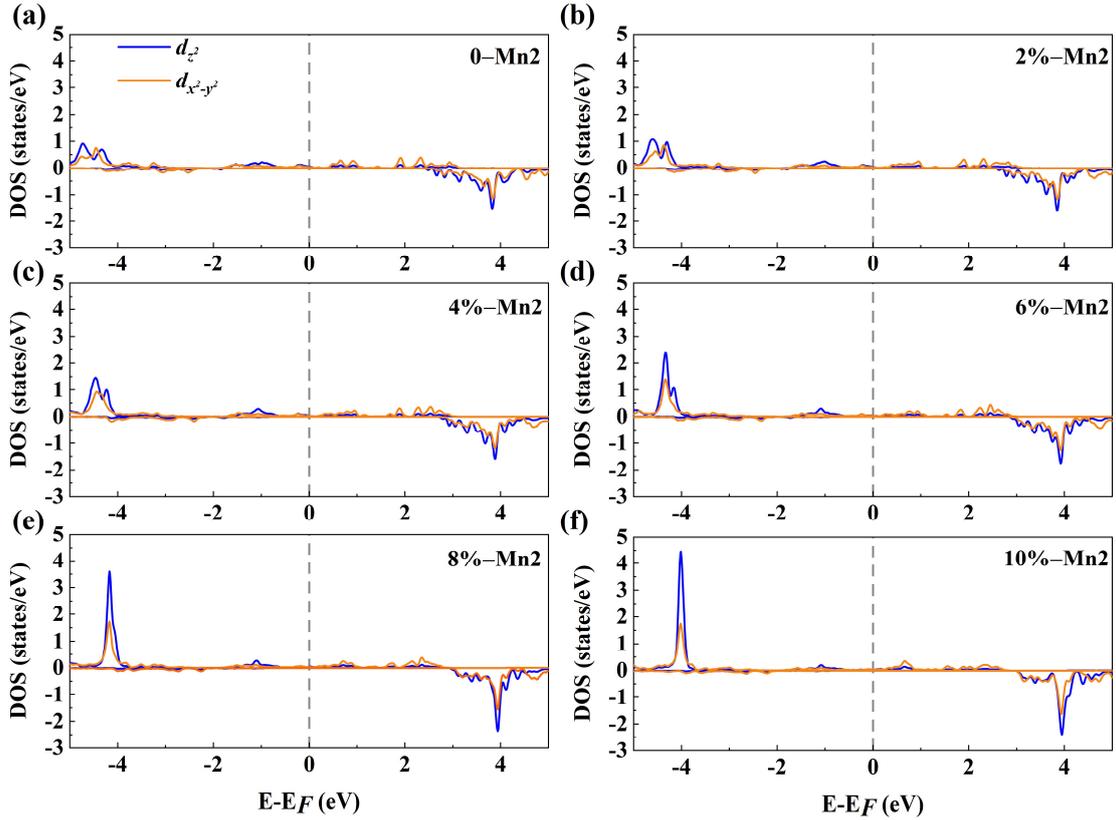

**FIG. 6.** The projected density of states (PDOS) on the Mn-3$d$ orbitals in MnO$_6$ octahedra (Mn2) at (a) 0 strain, biaxial tensile strain (b) 2%, (c) 4%, (d) 6%, (e) 8%, and (f) 10%. The gray perpendicular dashed lines denote the Fermi level.

## IV. CONCLUSIONS

In conclusion, we theoretically investigated the structural, magnetic, and electronic properties of oxygen-substituted NaMnF$_3$ under biaxial tensile strain by the first-principles approaches. We selected the simpler structure containing an oxygen octahedron to explore the underlying physical mechanism, which demonstrates that the introduced oxygen octahedra induce a transition from the intrinsic G-AFM NaMnF$_3$ to the high-temperature FM NaMnF$_{1.5}$O$_{1.5}$. Moreover, the interesting transition from an insulator to a half-metal makes it possible in spintronic applications. By applying the in-plane biaxial tensile strain, the $T_C$ increases significantly, and a room temperature $T_C$ may be expected. The changing trends of exchange coupling constants with the increasing of biaxial tensile strain can be attributed to the cooperative effects of Jahn-Teller distortion and rotation distortion. Our work may provide a new angle in designing novel high-temperature ferromagnets for the future theoretical and experimental works, and the coexisting half-metallic behavior suggests potential application in new spintronic devices.

## ACKNOWLEDGMENTS

This work is supported by the National Key Basic Research Program of China (Grant Nos. 2019YFA0308500 and 2020YFA0309100), the

National Natural Science Foundation of China (Grant Nos. 11721404, 51761145104, 11974390, and 11674385), the Key Research Program of Frontier Sciences of the Chinese Academy of Sciences (Grant No. QYZDJ-SSW-SLH020), the Youth Innovation Promotion Association of the Chinese Academy of Sciences (Grant No. 2018008), the Beijing Nova Program of Science and Technology (Grant No. Z191100001119112), Beijing Natural Science Foundation (2202060), and the Strategic Priority Research Program (B) of the Chinese Academy of Sciences (Grant No. XDB33030200).